\documentclass[11pt]{article}
\usepackage[utf8]{inputenc}

\usepackage{xspace}
\xspaceaddexceptions{\%}
\xspaceremoveexception{-}

\usepackage{authblk}
\usepackage{graphicx}
\usepackage{booktabs}

\usepackage[most]{tcolorbox}

\usepackage[inline]{enumitem}

\usepackage[margin=1in]{geometry}
\usepackage[final]{microtype}

\usepackage{amsmath,amsfonts,amssymb,amsthm}
\usepackage{braket}  %

\usepackage[backend=biber,style=ieee,maxcitenames=1,mincitenames=1]{biblatex}
\usepackage{xurl}
\usepackage[colorlinks]{hyperref}
\usepackage{cleveref}
\usepackage{xcolor} %
\def\tmp#1#2#3{%
  \definecolor{Hy#1color}{#2}{#3}%
  \hypersetup{#1color=Hy#1color}}
\tmp{link}{HTML}{800006}
\tmp{cite}{HTML}{2E7E2A}
\tmp{file}{HTML}{131877}
\tmp{url} {HTML}{8A0087}
\tmp{menu}{HTML}{727500}
\tmp{run} {HTML}{137776}
\def\tmp#1#2{%
  \colorlet{Hy#1bordercolor}{Hy#1color#2}%
  \hypersetup{#1bordercolor=Hy#1bordercolor}}
\tmp{link}{!60!white}
\tmp{cite}{!60!white}
\tmp{file}{!60!white}
\tmp{url} {!60!white}
\tmp{menu}{!60!white}
\tmp{run} {!60!white}

\newtheorem{theorem}{Theorem}[section]

\theoremstyle{definition}
\newtheorem{example}[theorem]{Example}

\DeclareMathOperator{\Tr}{Tr}

\newcommand{\alphasig}{\alpha}           %
\newcommand{\betatypetwo}{\beta}            %
\newcommand{\Pe}{P_{\mathrm{e}}}         %
\newcommand{\regimefactor}{\kappa}           %

\begin{document}

\title{The Cost of Certainty: Shot Budgets in Quantum Program Testing}
\author[]{Andriy Miranskyy}
\affil[]{Department of Computer Science, Toronto Metropolitan University \\ Toronto, Canada}
\affil[]{avm@torontomu.ca}

\date{}
\maketitle
\begin{abstract}
As quantum computing advances toward early fault-tolerant machines, testing and verification of quantum programs become urgent but costly, since each execution consumes scarce hardware resources. Unlike in classical software testing, every measurement must be carefully budgeted.

This paper develops a unified framework for reasoning about how many measurements are required to verify quantum programs. The goal is to connect theoretical error bounds with concrete test strategies and to extend the analysis from individual tests to full program-level verification.

We analyze the relationship between error probability, fidelity, trace distance, and the quantum Chernoff bound to establish fundamental shot count limits. These foundations are applied to three representative testing methods: the inverse test, the swap test, and the chi-square test. Both idealized and noisy devices are considered. We also introduce a program-level budgeting approach that allocates verification effort across multiple subroutines.

The inverse test is the most measurement efficient, the swap test requires about twice as many shots, and the chi-square test is easiest to implement but often needs orders of magnitude more measurements. In the presence of noise, calibrated baselines may increase measurement requirements beyond theoretical estimates. At the program level, distributing a global fidelity target across many fine-grained functions can cause verification costs to grow rapidly, whereas coarser decompositions or weighted allocations remain more practical.

The framework clarifies trade-offs among different testing strategies, noise handling, and program decomposition. It provides practical guidance for budgeting measurement shots in quantum program testing, helping practitioners balance rigour against cost when designing verification strategies.
\end{abstract}

\section{Introduction}

Quantum computing is approaching a transition: from noisy intermediate-scale quantum devices toward early fault-tolerant quantum computers~\cite{lanes2025framework,mandelbaum2025ibm,quantinuum2025quantinuum}. These advances will unlock applications well beyond the reach of simulators, but they also make testing and verification urgent and costly problems~\cite{miranskyy2019testing,miranskyy2020quantum,miranskyy2021testing,murillo2024challenges,ramalho2024testing}. While classical tests can often be executed at relatively low cost, each quantum program run consumes valuable hardware resources. Every measurement (or shot) must therefore be budgeted carefully, balancing statistical rigour against practical cost~\cite{mohammad2024meta,shonan2025qse}.

When implementing a quantum program, developers typically aim for a specific target state. In practice, however, the realized state may deviate due to code defects, errors introduced during compilation and transpilation, or imperfections on the device. While such discrepancies are easily detectable for small circuits, e.g., via direct state vector comparison~\cite{miranskyy2025feasibility, ye2025measurement}, this approach, in general, becomes infeasible because the classical resources needed to represent and compare state vectors grow exponentially with the number of qubits~\cite{miranskyy2025feasibility}. The key practical question is thus: \textit{How many shots are required to distinguish the actual and expected states with high confidence?} The number of shots represents a trade-off between quantity and quality. Although taking more measurements may seem to improve confidence, doing so can quickly deplete limited hardware resources. The goal is to collect enough data to obtain meaningful results without exhausting the measurement budget.

Prior work in quantum software engineering has approached this challenge from complementary perspectives. One study empirically compares the applicability of statevector-based validation (when feasible) with measurement-based methods such as inverse, swap, and statistical tests~\cite{miranskyy2025feasibility}. Other work argues that relying solely on measurement outcomes may be insufficient, motivating new strategies for output validation in quantum program testing~\cite{ye2025measurement}. A statistical line of research further demonstrates how sampling-based methods can be leveraged to uncover latent program bugs~\cite{sato2025bug}. Related research in quantum verification and characterization echoes similar themes, e.g., exploring resource vs. accuracy trade-offs in cross-entropy benchmarking, randomized benchmarking, and quantum process tomography, see~\cite{eisert2020quantum} for review. Together, these efforts highlight the spectrum of approaches available to practitioners: from exact but memory-intensive state vector methods, to scalable but sampling-limited measurement-based tests. However, what remains missing is a unified framework for reasoning about shot budgets, one that rigorously connects theoretical distinguishability bounds to concrete testing strategies under realistic hardware constraints.

In this work, we develop such a framework. At the theoretical level, we analyze how the quantum Chernoff bound (QCB)~\cite{audenaert2007discriminating,nussbaum2009chernoff}, fidelity~\cite{uhlmann1976transition,jozsa1994fidelity}, and trace distance~\cite[Sec. 9.2.1]{nielsen_chuang_2010} govern the number of measurements required to separate actual from expected states across pure-pure, pure-mixed, and mixed-mixed regimes. At the practical level, we evaluate three representative testing procedures~\cite{miranskyy2025feasibility}:
\begin{itemize}
    \item the inverse test, which directly overlaps actual and expected states;
    \item the swap test, which encodes fidelity through an ancillary qubit; and
    \item the chi-square test, which compares observed versus expected measurement distributions.
\end{itemize}

Our analysis spans both idealized and noisy conditions. Results show that the inverse test is the most sample-efficient, the swap test incurs roughly a factor-of-two overhead, and chi-square tests (while simple to implement) may require orders of magnitude more shots.

Beyond individual tests, we extend the analysis to the program level, where an application consists of multiple subroutines. We introduce the Bures angle~\cite{uhlmann1995geometric}, \cite[Eq. 9.32]{bengtsson2006geometry} as a natural tool for decomposing a global fidelity goal into per-function tolerances. This reveals a scaling challenge: verification costs grow rapidly when programs are decomposed into too many fine-grained functions, analogous to reliability engineering where overall system constraints tighten as more components are added in sequence.

Our contributions are as follows.
\begin{enumerate}
    \item Establishing theoretical shot-count bounds using QCB, fidelity, and trace distance, clarifying their behaviour across pure and mixed regimes;
    \item Deriving shot estimates for inverse, swap, and chi-square tests, with explicit trade-offs in efficiency and susceptibility to noise;
    \item Introducing a program-level budgeting framework based on the Bures angle, enabling systematic allocation of verification resources across program components; and
    \item Providing an interactive demonstration, available at \url{https://github.com/miranska/qse-shot-budget}.
\end{enumerate}
Together, these results provide both theoretical insight and practical guidance for budgeting measurement shots in quantum program testing, helping practitioners design verification strategies that are rigorous, scalable, and cost-aware.

The remainder of the paper is organized as follows. \Cref{sec:theory} develops the theoretical foundations linking QCB, fidelity, and trace distance to shot requirements. \Cref{sec:tests} applies these foundations to the inverse, swap, and chi-square tests, while \Cref{sec:noise_measurements} extends the analysis to noisy devices. \Cref{sec:error_budgeting} introduces program-level budgeting via the Bures angle and illustrates its use with examples. \Cref{sec:discussion} discusses implications, limitations, and avenues for future work, and \Cref{sec:conclusion} concludes.

\section{Theoretical foundations for shot estimation}\label{sec:theory}
Before turning to specific test procedures, we first establish theoretical foundations for estimating the number of measurement shots required to distinguish an actual quantum state from its expected counterpart. This section develops the relationships between error probability, fidelity, trace distance, and the QCB, which together provide a quantitative framework for shot budgeting. These tools map desired error tolerances into explicit shot estimates, under varying assumptions about whether the compared states are pure or mixed. \Cref{sec:qcb} introduces the QCB, \Cref{sec:fidelity} explores fidelity-based estimates, and \Cref{sec:trace_distance} provides an alternative formulation in terms of trace distance.

While we present formulations in terms of both fidelity and trace distance, in the remainder of the paper, we focus our analysis on fidelity. This choice streamlines the exposition, since all results can be reformulated in terms of trace distance by following the same derivation steps. Readers who prefer to think in terms of trace distance may therefore reinterpret the subsequent fidelity-based analysis accordingly.

\subsection{Quantum Chernoff bound}\label{sec:qcb}
Let $\rho$ and $\sigma$ be the density matrices of the \emph{actual} and \emph{expected} states.
After performing $N$ measurements (shots), the error probability $\Pe$ in distinguishing $\rho$ and $\sigma$ satisfies the QCB~\cite{audenaert2007discriminating,nussbaum2009chernoff}:
\begin{equation} \label{eq:_prob_estimate}
\Pe \sim e^{-N\xi_{\mathrm{QCB}}},
\quad
\xi_{\text{QCB}} = \lim_{N \to \infty} - \frac{\ln \Pe}{N} = - \ln Q(\rho,\sigma),
\quad
Q(\rho,\sigma) := \min_{0\le s\le 1}\Tr\left(\rho^s \sigma^{1-s}\right),
\end{equation}
where ``$\Tr$'' denotes the matrix trace.
Solving~\Cref{eq:_prob_estimate} for $N$ gives the asymptotic shot distance needed to achieve a target $\Pe\in(0,1)$:
\begin{equation}\label{eq:shot_estimate}
N \sim \frac{\ln \Pe}{\ln Q(\rho,\sigma)}.
\end{equation}
Although \Cref{eq:shot_estimate} is asymptotic in $N\to\infty$, an empirical study shows that it
remains accurate even for modest $N$~\cite{miranskyy2025feasibility}.

\Cref{{eq:shot_estimate}} is useful when both states are known in advance, e.g., 
\begin{enumerate*}[label=\roman*)]
    \item when evaluating whether a defect detector in the code works correctly or
    \item when checking that an original and a transpiled circuit are equivalent~\cite{miranskyy2025feasibility}. 
\end{enumerate*}
However, in practice, developers rarely know the precise nature of a defect during early debugging, nor the magnitude of deviations from the intended state.

\subsection{Fidelity}\label{sec:fidelity}
Instead, a more practical question is: \textit{Given an expected state $\sigma$, can we bound the error
probability so that the implemented state remains within a specified tolerance of the ideal state?} 
This is analogous to classical numerical analysis, where floating-point results are accepted as
correct if they fall within a tolerance. Quantum computing adopts the same principle. Here, the relevant closeness measure is often fidelity~\cite{jozsa1994fidelity,nielsen_chuang_2010}: if the realized state exceeds a fidelity threshold relative to the target, it is deemed acceptable~\cite{tsai2025benchmarking}.

Let us show how to connect the fidelity requirements to the number of measurement shots. This mapping allows us to determine the number of shots required to achieve a given fidelity level, or, conversely, to translate a fidelity tolerance into a shot budget.

The Uhlmann fidelity~\cite{uhlmann1976transition,jozsa1994fidelity} quantifies the similarity between two quantum states and is defined as
\begin{equation}\label{eq:fidelity}
F(\rho, \sigma) 
= \left[ \Tr \left( \sqrt{\sqrt{\rho} \sigma \sqrt{\rho}} \right) \right]^2
= \left(\left\|\rho^{1/2}\sigma^{1/2}\right\|_{1}\right)^{2},
\quad
F(\rho, \sigma) \in [0, 1],
\end{equation}
where $\|A\|_1 = \Tr(\sqrt{A^\dagger A})$ is the trace norm and ``$\dagger$'' denotes complex conjugate transpose. 
Fidelity $F=0$ indicates orthogonal states, while $F=1$ indicates identical states.
Thus, for $F(\rho, \sigma)$,  larger values indicate greater similarity.

\paragraph{Cases where $\rho$ and $\sigma$ are pure or one is mixed}
If at least one of $\rho$ or $\sigma$  is pure, the relationship between fidelity and $Q(\rho, \sigma)$ is simple. As shown in~\cite[p. 160501-4]{audenaert2007discriminating}, \cite{kargin2005chernoff}, \cite[p. 014302-2]{boca2009quantum},
\begin{equation}\label{eq:q_pure}
    Q(\rho, \sigma) = F(\rho, \sigma) = \Tr(\rho \sigma).
\end{equation}
Let us denote the number of shots by $N_\text{pure}$ when both states are pure and by $N_\text{pure-mixed}$ when only one state is pure. Substituting \Cref{eq:q_pure} into \Cref{eq:shot_estimate} gives the number of shots in these cases:
\begin{equation}\label{eq:shot_estimate_pure}
N_\text{pure} = N_\text{pure-mixed} \sim \frac{\ln \Pe}{\ln F(\rho, \sigma)}.
\end{equation}
In both cases, the number of shots follows the same functional form.

\paragraph{Both states $\rho$ and $\sigma$ are mixed case}
When both $\rho$ or $\sigma$ are mixed, $Q(\rho, \sigma)$ cannot be expressed exactly in terms of fidelity, but can be bounded as follows:
\begin{equation}\label{eq:q_mixed}
 1 - \sqrt{1-F(\rho,\sigma)}
\leq 
Q(\rho, \sigma) 
\leq 
\sqrt{F(\rho, \sigma)},
\end{equation}
see \Cref{sec:q_mixed_boundaries} for details.
Substituting \Cref{eq:q_mixed} into \Cref{eq:shot_estimate},  gives bounds on the required shot count (denoted by $N_\text{mixed}$):
\begin{equation}\label{eq:shot_estimate_mixed}
\frac{\ln \Pe}{\ln \left[1 - \sqrt{1-F(\rho,\sigma)}\right]}
\lesssim
N_\text{mixed}
\lesssim
\frac{\ln \Pe}{\ln \sqrt{F(\rho, \sigma)}}=\frac{2\ln \Pe}{\ln F(\rho, \sigma)}.
\end{equation}
Here, the symbol $\lesssim$ reflects the asymptotic ``$\sim$'' in \Cref{eq:shot_estimate}.

\subsubsection{Comparison of $N_\text{pure}$, $N_\text{pure-mixed}$, and $N_\text{mixed}$}
The behaviour of \Cref{eq:shot_estimate_pure,eq:shot_estimate_mixed} becomes clearer at the
extremes. 
For identical states ($F=1$), no finite number of shots suffices as $N_{\text{pure}} = N_{\text{pure-mixed}} = N_{\text{mixed}} \to \infty$. 
For orthogonal states ($F=0$), $N_{\text{pure}} = N_{\text{pure-mixed}} = N_{\text{mixed}} = 0$, meaning that a single shot is enough.  

\begin{figure}[tbh]
    \centering
    \includegraphics[width=1.0\linewidth]{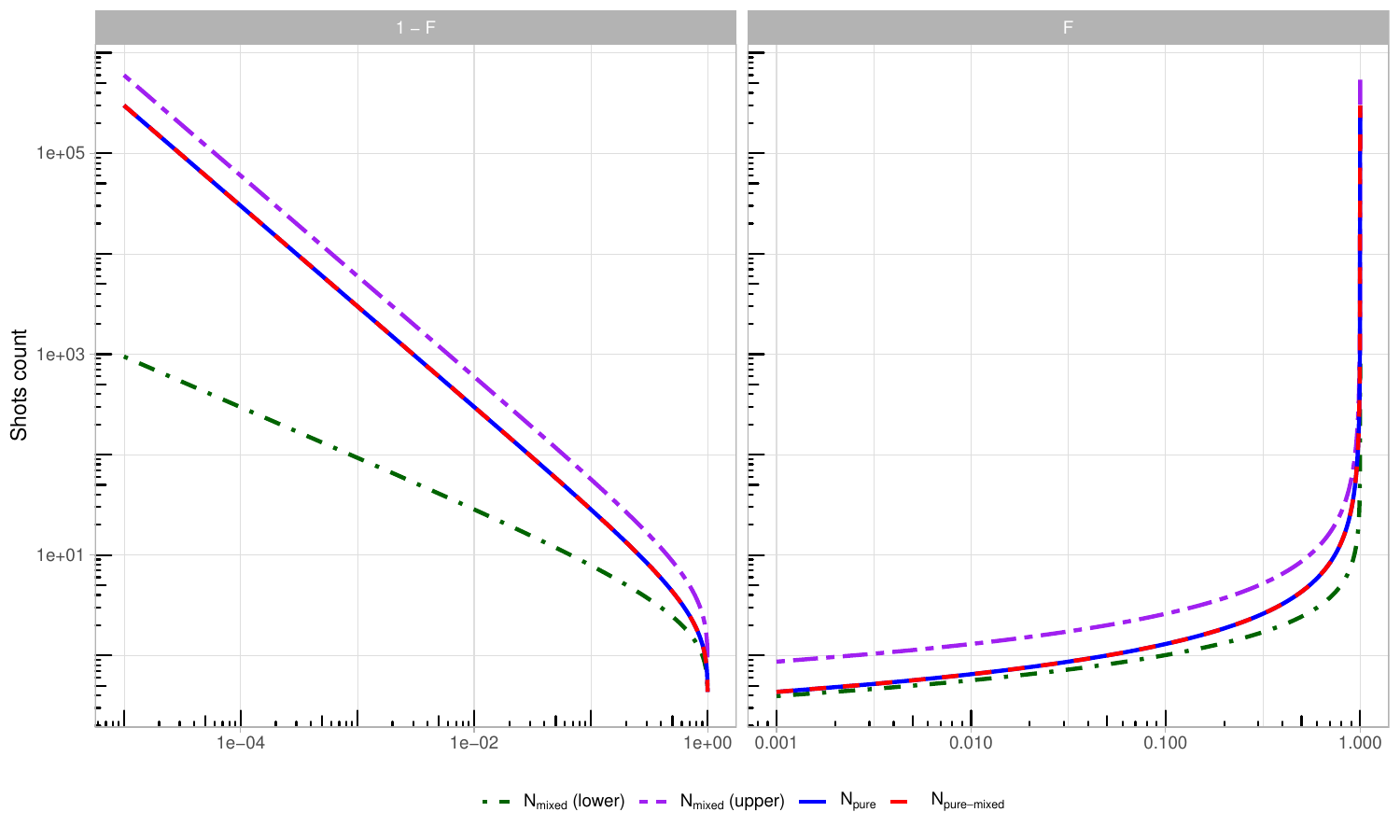}
    \caption{Number of measurement shots $N$ required to achieve error probability $\Pe = 0.05$ as a function of fidelity $F \in [0.001, 0.99999]$ (right pane).Curves are shown for the pure (or pure-mixed) case $N_{\text{pure}} = N_{\text{pure-mixed}}$ and for the lower and upper bounds of the mixed-mixed case $N_{\text{mixed}}$. As $F \to 0$, a single shot suffices; as $F \to 1$, the required shots diverge exponentially. To improve readability near $F=1$, the left pane re-expresses the data as a function of $1-F$, which makes the divergence more apparent.}
    \label{fig:f_vs_n}
\end{figure}

\Cref{fig:f_vs_n} plots the number of shots as a function of $F$. As $F \to 0$, the shot count approaches one, while as $F \to 1$, the required $N$ grows exponentially.
For cases with at least one pure state, $N_{\text{pure}}$ lies between the bounds of $N_{\text{mixed}}$.
Notably, the upper bound for $N_{\text{mixed}}$ is roughly twice $N_{\text{pure}}$ (or $N_{\text{pure-mixed}}$), while the lower bound can be much smaller.

\paragraph{} Having established these theoretical foundations, we next translate them into concrete test procedures (inverse, swap, and chi-square tests) and derive the corresponding shot estimates in Section~\ref{sec:tests}.

\section{Practical shot estimates: inverse, swap, and chi-square tests}\label{sec:tests}
Building on the theoretical foundations of \Cref{sec:theory}, we now turn to practical test procedures. We analyze three representative approaches: the inverse test (\Cref{sec:inverse_test}), which overlaps the actual and expected states directly; the swap test (\Cref{sec:swap_test}), which encodes fidelity through an ancillary qubit; and the chi-square test (\Cref{sec:chisq_test}), a statistical method comparing observed and expected measurement distributions.

For the inverse and swap tests, we derive closed-form shot estimates under both ideal and noisy conditions. For the chi-square test, our present treatment is restricted to the ideal regime, since extending distribution-based hypothesis testing to noisy devices requires more elaborate noise models. \Cref{sec:test_comparison} then compares the relative efficiency of the three approaches.

Throughout, we present results in terms of fidelity for clarity and brevity. However, all the derivations can be reformulated in terms of trace distance (see \Cref{sec:trace_distance}), so readers preferring that measure can map the results accordingly.

These analyses provide a practical foundation for choosing among the tests, highlighting trade-offs between circuit complexity, susceptibility to errors, and sampling overhead.

\subsection{Inverse test}\label{sec:inverse_test}
The inverse test (as described in~\cite[Sec. 3-B4]{miranskyy2025feasibility}) proceeds as follows. We first execute the actual circuit to obtain the state $\ket{\psi_A}$. We then apply the complex conjugate transpose of the expected state, $\ket{\psi_E}^\dagger$, and measure in the computational basis. 

If the actual and expected states are identical, i.e., $\ket{\psi_A} = \ket{\psi_E} $, then by construction the resulting state is $$
\ket{0^n} := \ket{0^{\otimes n}} = \ket{0_1 0_2 \dots 0_n},
$$
where $n$ is the width of the quantum register\footnote{Analogously, $\bra{0^n} := \bra{0^{\otimes n}} = \bra{0_1 0_2 \dots 0_n}$.}. In this case, every measurement produces the all-zero bitstring $(0_1 0_2 \cdots 0_n)$ of length $n$, denoted by $0^n$. 

If the states are only close, i.e., $\ket{\psi_A} \approx \ket{\psi_E} $, then it may take many shots before a nonzero bitstring is observed, indicating a deviation between the states. 
Formally, the probability of measuring the all-zero outcome equals the fidelity between the two pure (expected and actual) states:
$$
P(M_{\ket{\psi_R}} = 0^n) = \left | \braket{\psi_E | \psi_A} \right |^2,    
$$
where $M_{\ket{\psi_R}}$ denotes the measurement outcome of the quantum register at the end of the inverse test; see Appendix~\ref{sec:inverse_test_zero_string_probability} for details.
For pure states, the fidelity reduces to this squared inner product.

\subsubsection{Ideal quantum computer}\label{sec:inverse_ideal_qc}
On an ideal device, the expected outcome state is pure, because $\sigma = \ket{0^n} \bra{0^n}$. Substituting $Q(\rho, \sigma) = F(\rho, \sigma)$ into \Cref{eq:shot_estimate_pure}, the required number of shots is\footnote{\label{fn:inverse_probailistic}
Notably, we can reach a similar result with simple probabilistic reasoning. Let $F$ be the probability of observing a zero string; then the probability of failure is $1 - F$. We accept the test only if every trial yields a zero-string. With $N$ trials, the acceptance probability is $F^N$. To make the false acceptance probability $\le \Pe$, set
$$
    F^N \le \Pe \quad \Rightarrow \quad N \ge \frac{\ln(\Pe)}{\ln(F)},
$$
which is structurally similar to \Cref{eq:N_inverse_ideal}; however, note that the direction of the inequality differs. 
In this sense, the probabilistic argument is more pessimistic, since it demands that $N$ exceed this bound rather than treating it as an asymptotic estimate.
}
\begin{equation}\label{eq:N_inverse_ideal}
N_{\text{inverse, ideal}} 
\lesssim
\frac{\ln(\Pe)}{\ln F(\rho, \sigma)},
\end{equation}
where $\Pe$ specifies the tolerable error probability and $F(\rho,\sigma)$~---~the desired fidelity threshold. We will discuss setting specific values of $F$ in \Cref{sec:error_budgeting}.

\example[Inverse test at $F=0.999$]{\label{ex:inverse_ideal_0.999}
Suppose that we would like to make sure that our actual state is close to the expected states at $F(\rho, \sigma) = 0.999$ and we would like to have high confidence in our certainty, and thus we set $\Pe = 0.01$. Then, as per \Cref{eq:N_inverse_ideal}
$$
    N_{\text{inverse, ideal}} \lesssim \frac{\ln(\Pe)}{\ln F(\rho, \sigma)} = \frac{\ln(0.01)}{\ln(0.999)}  \approx 4603.
$$
}

\example[Inverse test at $F=0.99$]{\label{ex:inverse_ideal_0.99}
Now let us suppose that our practical use case suggests that we can relax our constraints and we are comfortable with $F(\rho, \sigma) = 0.99$; then
\begin{equation*}
    N_{\text{inverse, ideal}} \lesssim \frac{\ln(\Pe)}{\ln F(\rho, \sigma)} = \frac{\ln(0.01)}{\ln(0.99)}  \approx 458.
\end{equation*}
}

\subsubsection{Real quantum computer}\label{sec:inverse_real}
Even fault-tolerant quantum computers (especially the early ones~\cite{quantinuum2024quantinuum,dasu2025breakingmagicdemonstrationhighfidelity,daguerre2025experimental,lacroix2025scaling,daguerre2025experimental,dasu2025breakingmagicdemonstrationhighfidelity,goto2024high,bravyi2024high}) will have nonzero error rates associated with execution. For example, Quantinuum's first fault-tolerant quantum computer, expected to be delivered in 2029, is projected to achieve a logical error rate between $10^{-5}$ and $10^{-10}$~\cite{quantinuum2024quantinuum,dasu2025breakingmagicdemonstrationhighfidelity,daguerre2025experimental,lacroix2025scaling,daguerre2025experimental,dasu2025breakingmagicdemonstrationhighfidelity,goto2024high,bravyi2024high}.  In the future, the company aims to reduce this rate to $10^{-14}$~\cite{dasu2025breaking,quantinuum2025quantinuum}.

In this case, the expected state can be phenomenologically modelled as
$$
    \sigma = p_f (\ket{0^n} \bra{0^n}) + (1-p_f) \rho_\text{noise},
$$
where $p_f$ is the probability of obtaining the correct all-zero outcome after applying the inverse circuit, and $\rho_{\text{noise}}$ represents the residual weight spread across other computational basis states due to errors.

When $p_f=1$, we recover the ideal scenario of \Cref{sec:inverse_ideal_qc}. For a real device, $p_f < 1$ and the deviation of $p_f$ from unity capture the accumulated effect of gate errors, decoherence, measurement noise, and other imperfections. Thus, when analyzing test outcomes, $p_f$ serves as an ``effective survival probability'': it quantifies the chance of observing the all-zero outcome after applying the inverse circuit. 

This means that, when $p_f<1$, both the actual and the expected states are effectively mixed. The shot count estimate is then bounded between the pure, \Cref{eq:shot_estimate_pure}, and mixed-mixed regimes \Cref{eq:shot_estimate_mixed}:
$$
\frac{\ln \Pe}{ \ln F(\rho, \sigma)}
\lesssim
N_\text{inverse, real}
\lesssim
\frac{2 \ln \Pe}{ \ln F(\rho, \sigma)}.
$$
Equivalently, we may write
\begin{equation}\label{eq:shot_estimate_real_qc}
N_\text{inverse, real}
\lesssim
\frac{\regimefactor \ln \Pe}{ \ln F(\rho, \sigma)},
\end{equation}
where $\regimefactor \in [1,2]$ reflects whether one is in the pure or pure-mixed regime ($\regimefactor=1$) or the mixed-mixed worst case ($\regimefactor=2$).

The interval $\regimefactor \in [1,2]$ should be read as a sliding scale: when $p_f \to 1$, real devices tend to behave close to the pure-state estimate, while as noise grows or error channels misalign with the test, the cost drifts toward the upper bound. Thus, doubling the shot count is not always necessary, but serves as a conservative upper limit. In practice, more than doubling the shot count can happen due to various real-world imperfections; we will revisit this topic in \Cref{sec:noise_measurements}.

\subsection{Swap test}\label{sec:swap_test}
The version of the swap test discussed below is defined in~\cite[Sec. 3-B2]{miranskyy2025feasibility} and is based on the seminal swap test that is used to estimate the fidelity between two states~\cite{barenco1997,buhrman2001}. In this modified setup, the swap test functions as a binary detector rather than a fidelity estimator~---~execution continues until a nonzero outcome occurs or until sufficient confidence is achieved that the states are effectively identical.

The swap test provides an alternative to the inverse test: instead of executing the inverse circuit, the actual and expected states are compared indirectly using an ancillary qubit. The ancilla is first placed in superposition by a Hadamard gate, followed by a control-SWAP operation between the two registers, and then passed through a second Hadamard before being measured in the computational basis.

If the two states are identical, the ancilla ($q_a$) is always measured as $0$ (with probability $P = 1$). If the states are orthogonal, the probability of measuring $0$ drops to $0.5$ (see \cite[p.~167902-2]{buhrman2001} for details): 
\begin{equation}\label{eq:swap_p}
   P(M_{q_a} = 0) =  \frac{1}{2}+\frac{1}{2} F(\rho, \sigma),
\end{equation}
where $M_{q_a}$ denotes the measurement on the ancillary qubit $q_a$. Thus, unlike the inverse test, where orthogonal states are rejected with certainty, the swap test always retains a $0.5$ baseline acceptance probability (even for orthogonal states).

\subsubsection{Ideal quantum computer}\label{sec:swap_ideal}
We now estimate the number of shots required using the modified swap test. Unlike the inverse test, we cannot apply \Cref{eq:shot_estimate_pure} directly, since the zero string is no longer measured. Instead, the states are entangled with an ancilla, and the ancilla is measured. Fidelity still governs the outcome, but only through a shifted probability distribution. Thus, we do not observe fidelity itself, but a random variable whose expectation encodes it. To quantify this, we analyze the corresponding acceptance probability $Q$. In this case
$$
Q_{\text{swap}}(\rho, \sigma) = \frac{1}{2}+\frac{1}{2} F(\rho, \sigma),
$$
see \Cref{sec:qcb_for_swap} for details. By plugging this value into \Cref{eq:shot_estimate} we get\footnote{
We can also arrive at a similar answer using probabilistic reasoning.
The probability of not detecting any deviation after $N$ shots (i.e., obtaining $\ket{0}$ on every measurement) is
$$
P_{\text{miss}} = P(M_{q_a} = 0)^{N} = \left[\frac{1}{2}+\frac{1}{2} F(\rho, \sigma)\right]^{N}.
$$
To ensure that the probability of false acceptance does not exceed a chosen threshold $\Pe$, we require
$$
P_{\text{miss}} \le \Pe
\quad\Rightarrow\quad
N \ge \frac{\ln \Pe}{\ln\left(\frac{1+F}{2}\right)}.
$$
This expression is structurally similar to \Cref{eq:N_swap_ideal}, although note the difference in the direction of inequality. 
As in the inverse test case (\Cref{fn:inverse_probailistic}), the probabilistic argument is more pessimistic, since it enforces a lower bound on $N$ rather than providing an asymptotic estimate.
}
\begin{equation}\label{eq:N_swap_ideal}
N_{\text{swap, ideal}}
\lesssim
\frac{\ln(\Pe)}{\ln Q_{\text{swap}}} 
=
\frac{\ln(\Pe)}{\ln\left[ \frac{1}{2}+\frac{1}{2} F(\rho, \sigma) \right]}.
\end{equation}

To compare with the inverse test, we expand both \Cref{eq:N_inverse_ideal,eq:N_swap_ideal} around $F \to 1$ using Taylor series denoted by ``TS''. The ratio becomes
$$
\frac{N_{\text{swap, ideal}}}{N_{\text{inverse, ideal}}} = \frac{\ln F(\rho, \sigma)}{\ln\left[ \frac{1}{2}+\frac{1}{2} F(\rho, \sigma) \right]}
\overset{\text{TS at } F \to 1}{=}
2 - \frac{ F(\rho, \sigma)-1}{2} + O\left[(F(\rho, \sigma)-1)^2\right]
\approx
2,
$$
demonstrating that the swap test requires approximately twice as many shots as the inverse test. The extra cost arises because the fidelity is encoded indirectly via the ancilla rather than measured directly. The following two examples confirm this observation.

\example[Swap test at $F=0.999$]{
Suppose we target $F(\rho, \sigma) = 0.999$ with $P_e = 0.01$. Then, by \Cref{eq:N_swap_ideal},
$$
N_{\text{swap, ideal}}
\lesssim
\frac{\ln(\Pe)}{\ln\left[ \frac{1}{2}+\frac{1}{2} F(\rho, \sigma) \right]}
= 
\frac{\ln(0.01)}{\ln(0.9995)}
\approx 9208.
$$
which is approximately twice the $4603$ shots required by the inverse test at the same parameters (\Cref{ex:inverse_ideal_0.999}).
}

\example[Swap test at $F=0.99$]{
If we relax the constraint to $F(\rho, \sigma) = 0.99$ while keeping $P_e = 0.01$, then
$$
N_{\text{swap, ideal}}
\lesssim
\frac{\ln(\Pe)}{\ln\left[ \frac{1}{2}+\frac{1}{2} F(\rho, \sigma) \right]}
= 
\frac{\ln(0.01)}{\ln(0.995)}
\approx 919.
$$
roughly double the 458 shots required by the inverse test in \Cref{ex:inverse_ideal_0.99}.
}

\subsubsection{Real quantum computer}
As with the inverse test (\Cref{sec:inverse_real}), real devices exhibit noise from gate infidelities, decoherence, crosstalk, and measurement errors.
We therefore model the ``correct'' ancilla outcome $\ket{0}$ as occurring with probability $p_f$, with the residual distributed across other outcomes due to noise. When $p_f = 1$ we recover the ideal scenario (\Cref{sec:swap_ideal}); for $p_f < 1$, the effective states are mixed. 

In this mixed-mixed regime, the required shot count is bounded between the pure-state estimate and twice that amount:
\begin{equation}\label{eq:swap_real_bound}
N_{\mathrm{swap, real}} \lesssim \frac{\regimefactor \ln \Pe}{\ln\left[\frac{1}{2} + \frac{1}{2}F(\rho,\sigma)\right]},
\end{equation}
where (as in \Cref{sec:inverse_real}) $\regimefactor \in [1,2]$ interpolates between the pure or pure-mixed case ($\regimefactor=1$) and the conservative mixed-mixed worst case ($\regimefactor=2$). In practice, $\regimefactor$ can exceed~2 under severe noise. We will return to this issue in \Cref{sec:noise_measurements}.

Thus, while the swap test provides a practical alternative to the inverse test without requiring inverse circuit construction, it does so at the cost of approximately double the shot count in the ideal regime, with additional overhead possible under realistic noise.

\subsection{Chi-square test}\label{sec:chisq_test}

A third approach to evaluating whether an actual state deviates from the expected one is to employ statistical tests on the measurement distributions directly. In this setting, we do not attempt to reconstruct quantum state overlaps (as in the inverse or swap tests). Instead, we run the circuit under test, record its measurement outcomes in the computational basis, and compare the resulting empirical distribution $p=(p_1,\dots,p_k)$ against the theoretical expected distribution $q=(q_1,\dots,q_k)$. Classical hypothesis testing, in particular the chi-square goodness-of-fit test~\cite{pearson1900x}, \cite[pp. 45--52]{mcdonald2014handbook}, provides the statistical machinery for this comparison.

\subsubsection{Chi-square test formulation}
Let $N$ denote the total number of measurement shots, $O_i$ denote the observed counts for the $i$-th outcome, and $E_i = N q_i$ the expected counts under the theoretical distribution $q$. The empirical probabilities are $p_i = O_i / N$. The Pearson chi-square statistic~\cite{pearson1900x,mcdonald2014handbook} is defined as
$$
\chi^2 = \sum_{i=1}^k \frac{(O_i - E_i)^2}{E_i}.
$$
Under the null hypothesis that the actual distribution is equal to the expected one ($p=q$), $\chi^2$ approximately follows a chi-square distribution with $k-1$ degrees of freedom.

To ensure that the test detects a discrepancy with significance $\alphasig$ and power $1-\betatypetwo$, the required sample size $N$ can be obtained as follows:
\begin{enumerate}
\item Compute the effect size (also called the \textit{$\chi^2$-distance}~\cite[p. 425]{gibbs2002choosing}):
\begin{equation}\label{eq:chi2_distance}
    d_{\chi^2}(p, q) = w^2
= \sum_{i=1}^k \frac{( p_i - q_i)^2}{q_i}.
\end{equation}

\item Under the alternative hypothesis ($p \ne q$), the test statistic follows a noncentral chi-square distribution with noncentrality parameter $\lambda(k-1, \alphasig, 1-\betatypetwo) = N w^2$, see~\cite[Sec. 12.7.1]{cohen1988statistical} for details.

\item The value of $\lambda$ can be computed numerically; then $N$ is obtained~\cite[Sec. 12.7.2]{cohen1988statistical} by
\begin{equation}\label{eq:n_chisq}
N = \frac{\lambda(k-1, \alphasig, 1-\betatypetwo)}{w^2}.
\end{equation}
\end{enumerate}
Here, $\alphasig$ specifies the Type~I error rate (false positives) and $\betatypetwo$ the Type~II error rate (false negatives). Thus, $\alphasig$ and $\betatypetwo$ directly quantify the risks of over- and under-detecting meaningful deviations. By contrast, inverse and swap tests on ideal devices have zero false-positive probability by construction~\cite{miranskyy2025feasibility}, leaving only the analogue of $\betatypetwo$. In this sense, $\betatypetwo$ in the chi-square test plays the same role as $\Pe$ in those tests.

As with all chi-square tests, validity requires that expected counts not be too small. Standard heuristics are $E_i \geq 5$ for all bins~\cite[p. 420]{cochran1954some} and $N \geq 13$~\cite{pearson1900x}. For quantum circuits with sparse or peaked distributions, these conditions may fail, requiring alternative techniques such as resampling.

\subsubsection{Ideal quantum computer: connecting chi-square test to fidelity}
\paragraph{General analysis}
To compare with inverse and swap tests, we connect $w^2$ to fidelity. We proceed via the \textit{Hellinger distance}~\cite[p. 422]{gibbs2002choosing}:
$$
d_{H}(p, q)
= \left[ \frac{1}{2} \sum_{i=1}^{k} \big(\sqrt{p_i} - \sqrt{q_i}\big)^2 \right]^{1/2}
= \left[ 1 - \sum_{i=1}^{k} \sqrt{p_i q_i}\right]^{1/2}.
$$
The last equality uses normalization $\sum_{i=1}^{k} p_i = \sum_{i=1}^{k} q_i = 1$. The term $\sum_{i=1}^{k} \sqrt{p_i q_i}$ is the \textit{Bhattacharyya coefficient}~\cite{bhattacharyya1943measure,kailath1967divergence}, a measure of distributional overlap. 

The following bound holds~\cite[p. 429]{gibbs2002choosing}:
$$
d_{H}(p, q)  \le \sqrt{2} \left[d_{\chi^2}(p, q) \right]^{1/4}.
$$
Therefore,
\begin{equation}\label{eq:chi_vs_hellinger}
d_{\chi^2}(p, q)
\ge
\frac{1}{4}\left[d_{H}(p, q)\right]^4
=
\frac{1}{4}\left(1 - \sum_{i=1}^{k} \sqrt{p_i q_i}\right)^{2}.
\end{equation}
Moreover, since
\begin{equation}\label{eq:clssic_vs_quantum_fidelity}
0 \le \sqrt{F(\rho, \sigma)} \le \sum_{i=1}^{k} \sqrt{p_i q_i} \le 1,
\end{equation}
as per~\cite{luo2004informational}, \cite[Eq. 3.154]{watrous2018theory}), we have
$$
\left(1 - \sum_{i=1}^{k} \sqrt{p_i q_i}\right)^{2} 
\le 
\left(1 - \sqrt{F(\rho,\sigma)}\right)^{2}.
$$
Thus, without additional measurements assumptions, no tighter bound can be obtained solely in terms of $F(\rho,\sigma)$. The challenge arises because \Cref{eq:chi2_distance} is asymmetric and highly sensitive to small denominators, whereas smoother, symmetric distances (e.g., Hellinger or fidelity-based) behave more stably.

To build intuition despite these limitations, we next examine two specific use cases.

\paragraph{Specific case: fidelity-attaining measurement}
Suppose that the readout $E$ is chosen so that the classical overlap attains the quantum fidelity~\cite[Sec. 2]{fuchs1995mathematical}:
$$
F(\rho,\sigma) = \min_E \sum_i \sqrt{p_i q_i}
= \sum_i \sqrt{p_i q_i}.
$$
Then the inequality in \Cref{eq:clssic_vs_quantum_fidelity} tightens, and \Cref{eq:chi_vs_hellinger} gives
\begin{equation}\label{eq:w2_fidelity_attaining}
d_{\chi^2}(p, q) = w^2 \ge \frac{1}{4} \left(1-\sqrt{F(\rho,\sigma)}\right)^2.
\end{equation}
Substituting into \Cref{eq:n_chisq}, the lower bound for the number of shots required in this case satisfies
\begin{equation}\label{eq:N_fidelity_attaining}
N_{\chi^2\text{-attaining, ideal}}
\le \frac{\lambda(k-1, \alphasig, 1-\betatypetwo)}{w^2}
= \frac{\lambda(k-1, \alphasig, 1-\betatypetwo)}{\frac{1}{4}\left[1-\sqrt{F(\rho, \sigma)}\right]^2}.
\end{equation}

This represents an optimistic bound: if the observed classical overlap exceeds the quantum fidelity (as it often does in practice), then $w^2$ is smaller, and \Cref{eq:n_chisq} implies that many more shots are needed.

\paragraph{Specific case: small discrepancy}
Now consider a different regime: the actual and expected distributions are very close (i.e., the difference is subtle), and we want to detect a subtle deviation. In such a scenario $F \approx 1$. We can model it by supposing that 
$$
p_i = q_i + \delta_i,\qquad \sum_i \delta_i = 0,\qquad q_i>0,\qquad |\delta_i|\ll q_i.
$$
In this scenario $w^2$ becomes
$$
d_{\chi^2}(p, q) = w^2
= \sum_{i=1}^k \frac{( p_i - q_i)^2}{q_i} 
= \sum_{i=1}^k \frac{( q_i + \delta_i-q_i)^2}{q_i} 
= \sum_{i=1}^k \frac{\delta_i^2}{q_i}.
$$
Expanding the Hellinger distance for small $\delta_i$ gives
\begin{align*} 
d_{H}(p, q) 
&= \left[ \frac{1}{2} \sum_{i=1}^{k} \left(\sqrt{p_i} - \sqrt{q_i}\right)^2 \right]^{1/2}
= \left[ \frac{1}{2} \sum_{i=1}^{k} \left(\sqrt{q_i-\delta_i} - \sqrt{q_i}\right)^2 \right]^{1/2} \\
&\overset{\text{TS as } \delta_i \to 0}{=}
\left[ \frac{1}{2} \sum_{i=1}^{k} \left(\sqrt{q_i} - \frac{\delta_i}{2\sqrt{q_i}} + O\left(\delta^2\right) - \sqrt{q_i}\right)^2 \right]^{1/2}
\approx
\left[ \frac{1}{8} \sum_{i=1}^{k}  \frac{\delta_i^2}{q_i} \right]^{1/2}
.
\end{align*}
Combining with \Cref{eq:clssic_vs_quantum_fidelity}, we obtain
\begin{equation}\label{eq:w2_small_discr}
w^2 \approx 8 [d_{H}(p, q)]^2 
= 8 \left[ 1 - \sum_{i=1}^{k} \sqrt{p_i q_i}\right]
\le 
8  \left[ 1 - \sqrt{F(\rho, \sigma)} \right].
\end{equation}
\Cref{eq:w2_small_discr} says that, when $p$ and $q$ are very close,
Pearson’s effect size $w^2$ decreases with the increase of fidelity. Note that it provides an \emph{upper envelope}:
the actually realized $w^2$ may be (and often is, based on empirical results in~\cite{miranskyy2025feasibility}) smaller, depending on how well the chosen readout ``sees'' the discrepancy. 

By \Cref{eq:n_chisq}, the lower bound on the required shots is then
\begin{equation}\label{eq:n_small_discr}
N_{\chi^2\text{-small, ideal}} 
\ge \frac{\lambda(k-1, \alphasig, 1-\betatypetwo)}{w^2} 
= \frac{\lambda(k-1, \alphasig, 1-\betatypetwo)}{8  \left[ 1 - \sqrt{F(\rho, \sigma)} \right]}.
\end{equation}

Let us look at two examples. In all of them, we compute $\lambda(k, \alphasig, \betatypetwo)$ numerically using \texttt{pwr.chisq.test} function in R~v.4.4.1~\cite{r2024} \texttt{pwr}~v.1.3-0~\cite{champely2020pwr} package.

\begin{example}[Chi-square test at $F=0.999$]
\label{ex:chi_square_f999}
Suppose the target fidelity is $F = 0.999$, with $k = 16$ bins (giving $15$ degrees of freedom). 
For $\alphasig = \betatypetwo = 0.01$, the function \texttt{pwr.chisq.test} yields $\lambda(15, 0.01, 0.99) \approx 44.93$.

For the fidelity-attaining measurement case, using \Cref{eq:w2_fidelity_attaining}, we obtain $w^2 \approx 6.25 \times 10^{-8}$ and, from \Cref{eq:N_fidelity_attaining}, 
$N_{\chi^2\text{-attaining, ideal}} \le 7.18 \times 10^{8}$.
For the small discrepancy case, based on \Cref{eq:w2_small_discr}, $w^2 \approx 4.00 \times 10^{-3}$ and, from \Cref{eq:n_small_discr}, 
$N_{\chi^2\text{-small, ideal}} \ge 1.12 \times 10^{4}$.

Thus, even in these two special cases, the chi-square test requires between $\sim 1.12 \times 10^{4}$ and $\sim 7.18 \times 10^{8}$ shots, the latter being prohibitively expensive.
\end{example}

\begin{example}[Chi-square test at $F=0.99$]
\label{ex:chi_square_f99}
Now suppose the target fidelity is $F = 0.99$, while keeping $k$, $\alphasig$, and $\betatypetwo$ the same as in \Cref{ex:chi_square_f999}. 
We again obtain $\lambda(15, 0.01, 0.99) \approx 44.93$.

For the fidelity-attaining measurement case, using \Cref{eq:w2_fidelity_attaining}, we find $w^2 \approx 6.28 \times 10^{-6}$ and, from \Cref{eq:N_fidelity_attaining}, 
$N_{\chi^2\text{-attaining, ideal}} \le 7.15 \times 10^{6}$.
For the small discrepancy case, based on \Cref{eq:w2_small_discr}, $w^2 \approx 4.01 \times 10^{-2}$ and, from \Cref{eq:n_small_discr}, 
$N_{\chi^2\text{-small, ideal}} \ge 1.12 \times 10^{3}$.

Compared to \Cref{ex:chi_square_f999}, the range narrows considerably, from $\sim 1.12 \times 10^{3}$ to $\sim 7.15 \times 10^{6}$ shots, though the upper bound remains costly.
\end{example}

\subsubsection{Real quantum computer}\label{sec:chisq_real}

The chi-square analysis so far has assumed idealized measurement distributions. However, on a real device, noise alters the baseline statistics, and extending the framework requires modelling this baseline explicitly. In practice, state preparation and measurement errors, device drift, and correlated noise all contribute, making the effective ``null distribution'' different from the theoretical ideal.

A natural way forward is to calibrate a noisy baseline distribution that reflects the device's behaviour in the absence of true defects. This baseline may be estimated using quantum goodness-of-fit and optimal measurements, control circuits, mirror-circuit benchmarking, or drift-detection techniques~\cite{temme2015goodness,proctor2020detecting,proctor2022mirror}. Once established, the chi-square test can then be applied in a one-sided fashion, checking whether the empirical distribution deviates beyond the noise floor. This approach resembles cross-entropy benchmarking~\cite{arute2019supremacy,boixo2018character}, where observed outcomes are compared against calibrated reference distributions to detect systematic deviations.

Conceptually, one may reinterpret \Cref{eq:chi2_distance} with the expected distribution $q$ replaced by this calibrated baseline. The statistical guarantees of the classical chi-square framework then carry over, but with respect to the device-calibrated reference rather than the ideal distribution. In principle, this allows practitioners to test for meaningful deviations while tolerating the stochastic fluctuations induced by noise.

The challenge is that baseline estimation (especially on non-fault-tolerant devices) itself is resource-intensive and subject to drift, while correlated or time-dependent errors can obscure genuine discrepancies. Developing robust statistical procedures that can separate real defects from noise-induced variation therefore remains an open and important research direction.

Despite these challenges, baseline-driven chi-square testing has a practical advantage: it integrates naturally with existing quantum characterization, verification, and validation workflows. Rather than requiring new circuits, it builds on established benchmarking methods.

We return to the impact of noise on the budgeting of the shot in \Cref{sec:noise_measurements}, where both the inverse and the swap tests are analyzed in the presence of device noise.

\subsection{Comparison of methods}\label{sec:test_comparison}
The preceding analysis shows that the swap test consistently requires about twice as many shots as the inverse test, while the chi-square test typically demands much more.

\begin{figure}[tbh]
    \centering
    \includegraphics[width=0.8\linewidth]{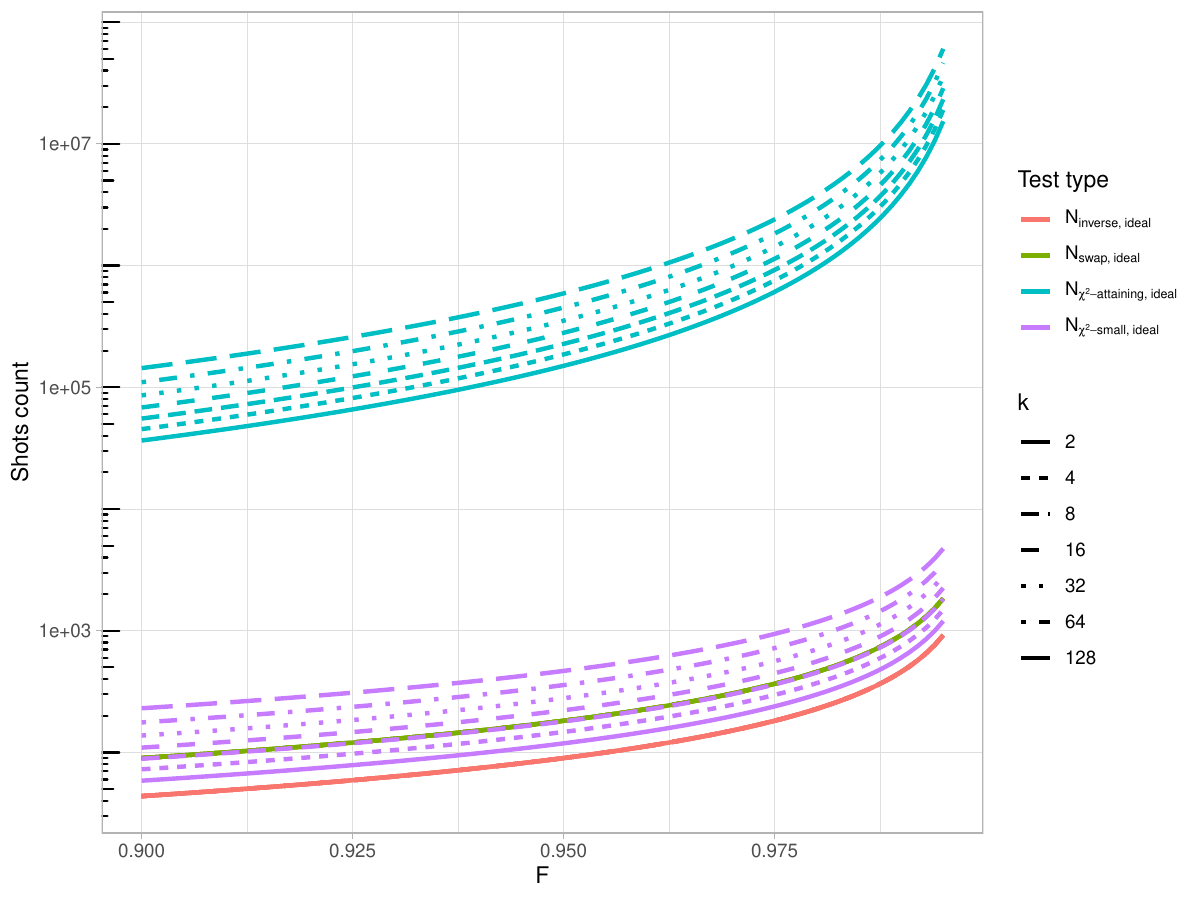}
    \caption{Required number of shots for inverse, swap, and chi-square tests as a function of fidelity $F \in [0.900, 0.995]$. Parameters are fixed at $\Pe = \alphasig = \betatypetwo = 0.01$. For the chi-square test, bin counts are varied over $k = 2, 4, 8, 16, 32, 64, 128$; inverse and swap tests are independent of $k$. The chi-square curves illustrate the wide range of possible sampling costs.}
    \label{fig:inverse_swap_chi}
\end{figure}

\Cref{fig:inverse_swap_chi}  illustrates these differences for fidelities\footnote{Fidelity is capped at $F = 0.995$ to maintain readability of \Cref{fig:inverse_swap_chi}. Beyond this point, the curves diverge rapidly, and the chi-square bounds in particular span several orders of magnitude, obscuring visual comparison.} $F \in [0.900, 0.995]$, with error parameters fixed at $\Pe = \alphasig = \betatypetwo = 0.01$. For the chi-square test, $k = 2, 4, 8, 16, 32, 64, 128$; whereas the inverse and swap tests remain independent of $k$.

The results reveal several trends. Increasing the number of bins raises the required shot count. In the small-discrepancy case, the chi-square test can yield shot counts comparable to inverse/swap when $k$ is small, but it already exceeds the swap test once $k \geq 16$.

In contrast, in the fidelity-attaining case, the chi-square test is orders of magnitude more demanding (by at least two orders when $F \approx 0.9$, and by nearly four orders as $F \to 0.995$). This divergence highlights the steep cost of high-fidelity verification with distribution-based tests.

Although the two chi-square scenarios shown represent only bounds within the possible range, they demonstrate that chi-square testing can be far more resource-intensive (in terms of the number of shots) than inverse or swap tests. Empirical results confirm this ordering~\cite[Figs.~7 and~8]{miranskyy2025feasibility}: inverse tests typically require the fewest shots, followed by swap tests, with chi-square tests demanding the most. Rarely, chi-square tests may yield lower shot counts due to stochastic variation, but such cases are exceptional.

\paragraph{Summary}
Inverse and swap tests provide direct fidelity-based shot estimates. The swap test typically incurs about a factor-of-two overhead compared to the inverse test, while the chi-square test often requires orders of magnitude more. On ideal devices, they are susceptible only to Type~II errors. Their main drawback is the need for circuit modifications (see~\cite[Tbl. 2]{miranskyy2025feasibility} for complexity analysis) and, for the swap test specifically, the expansion of the qubit register from $n$ to $2n+1$. 

By contrast, chi-square tests are cheaper to implement since they operate directly on measurement distributions, but this comes at the cost of much higher sample requirements~---~particularly as the number of bins grows\footnote{Because $w^2$ weights each term by $1/q_i$, very small $q_i$ values can inflate $w^2$. A common practical remedy is to coarse-grain (merge) bins so that all expected counts exceed standard thresholds (e.g., $E_i \ge 5$). This stabilizes $w^2$ and improves the validity of the $\chi^2$ approximation. In effect, scenarios can be designed where the effective number of bins satisfies $k \ll 2^n$.}. It can also be challenging to construct an accurate expected distribution for a noisy device. Moreover, chi-square tests are vulnerable to both Type~I and Type~II errors. In practice, then, the choice between these approaches requires balancing circuit complexity against sampling cost.

With this comparison established, we now examine how device noise alters these estimates in \Cref{sec:noise_measurements}.

\section{Impact of noise on shot estimates for inverse and swap tests}\label{sec:noise_measurements}
In \Cref{sec:tests}, we considered noise from a theoretical perspective. 
There, the QCB indicates that the required number of shots may need to be doubled, depending on whether the states are effectively pure or mixed. 
However, this adjustment only captures the regime change (pure versus mixed) and does not reflect the actual magnitude or structure of the noise generated by a device. 
As a result, the QCB-based estimates should be viewed as lower bounds: in practice, real devices emit noise of varying strength and type, and verification can demand substantially more shots.  

To address this gap, we now explicitly incorporate device noise into the analysis. 
The goal is to distinguish outcomes caused merely by random fluctuations from those that indicate genuine test failures. 
This turns verification into a statistical problem: additional repetitions are required to control both Type~I and Type~II errors while accounting for the device’s noise floor. 
We approach this by calibrating a noise-only baseline for both inverse and swap tests, and then computing the required number of shots as a function of the noise level and error tolerances.  

In this section, we focus exclusively on inverse and swap tests. These tests admit closed-form fidelity-based estimates that can be naturally extended to noisy settings. By contrast, adapting distribution-based tests such as the chi-square to noise requires complex calibrated baselines (\Cref{sec:chisq_real}), which we leave for future work.

There are various possible strategies for handling noise, depending on the desired precision and acceptable complexity (see~\cite{cai2023quantum} for a review). Here, we describe a simple method\footnote{Other methods, such as~\cite{virani2025distinguishing}, can be explored.}  inspired by error-analysis techniques in~\cite{proctor2020detecting,proctor2022scalable}. The key idea is to calibrate the process by constructing a calibrated noise baseline against which the results of the inverse or swap test are compared~\cite{proctor2022scalable}. This calibrated noise baseline distinguishes nonzero measurement outcomes caused by random hardware noise (such as readout errors or stochastic bit flips) from those indicating a real deviation between the expected and actual quantum states.

The calibration step involves running a control circuit that prepares and measures the all-zero state, but with gate depth and topology similar to the test circuit. This ensures that the baseline incorporates comparable noise effects. 
Several approaches, such as randomized compilations to identity~\cite{cai2023quantum}, can be used to construct such a circuit. 
The resulting all-zero probability, denoted $q_0$, provides an empirical estimate of the device's intrinsic noise floor. Deviations from this baseline in the actual inverse or swap test can then be interpreted as evidence of real state differences rather than random fluctuations.

The inverse test circuit is executed repeatedly, recording the number of all-zero outcomes $X$ out of $N$ total shots. The empirical probability $\hat{q} = X / N$ is compared to the calibrated noise baseline $q_0$ using a one-sided\footnote{A one-sided binomial test is used because only deficits relative to the calibrated baseline probability $q_0$ indicate a real discrepancy between the prepared and target states. An excess of ``correct'' outcomes simply reflects better-than-expected  performance.} binomial hypothesis test. Under the null hypothesis, the deviations arise solely from random noise at rate $1 - q_0$; under the alternative hypothesis, they exceed the noise-only rate, suggesting that the prepared and target states differ.

A similar logic would apply to the swap test: here, the baseline is the probability of measuring the ancilla qubit in state $\ket{0}$ under a control configuration. The observed ancilla outcome distribution in the actual swap test is then compared with this calibrated noise baseline through a binomial test. While the inverse and swap tests share the same noise-calibration framework, it is important to note that the swap test inherently requires about twice the number of shots as the inverse test (\Cref{sec:swap_test}). This factor-of-two overhead remains present under noise.

\example[Baseline calibration under noise]{\label{ex:noise_binomial}
Suppose we wish to distinguish the actual and expected states at target probability $q_1 \in \{0.90, 0.99\}$  (here $q_1$ plays the role of effective fidelity) and estimate how many shots are needed for a given baseline $q_0$. We compare these values with the recommendations $N_{\text{inverse, real}}$, \Cref{eq:shot_estimate_real_qc}, and $N_{\text{swap, real}}$, \Cref{eq:swap_real_bound}. Since the computations are performed on a noisy device, we set $\regimefactor = 2$ in those equations to represent the conservative mixed-mixed regime.

\Cref{fig:noise_binomial} shows the results. As expected, the closer the baseline is to the target, the more shots are required to distinguish them (e.g., when $q_0 = 0.991$ and $q_1 = 0.99$, the small gap of only $0.001$ inflates the required number of shots by more than two orders of magnitude).

Notably, even when $q_0 = 1.0$ (representing an ideal device), the binomial approach recommends more shots than the QCB-based estimates. This occurs because the binomial framework explicitly controls both Type~I and Type~II errors: even tiny deviations from perfect outcomes must be distinguished from random fluctuations, which require additional repetitions. The example thus illustrates that noise-aware calibration can make verification substantially more demanding than suggested by QCB alone.

\begin{figure}[tbh]
    \centering
    \includegraphics[width=0.7\linewidth]{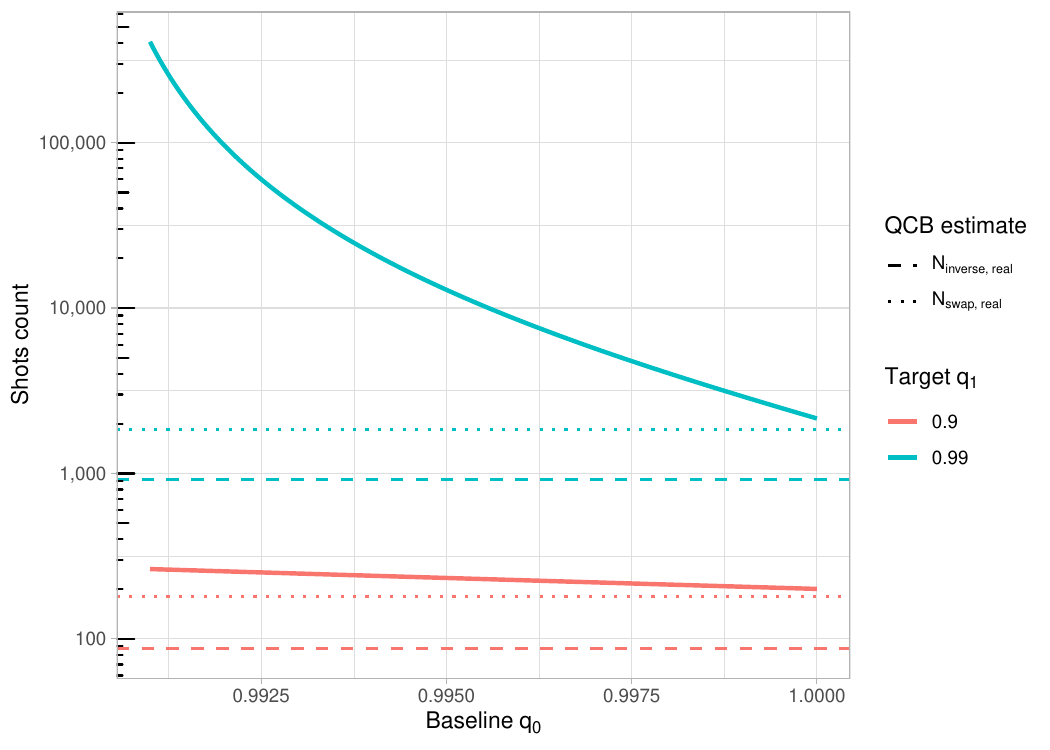}
    \caption{
    Shot-count requirements for \Cref{ex:noise_binomial}, comparing binomial-based estimates with QCB-based estimates $N_{\text{inverse, real}}$ and $N_{\text{swap, real}}$ with $\Pe=0.01$ and $\regimefactor = 2$. The target probabilities are $q_1 \in \{0.90, 0.99\}$, while the calibrated noise baseline is $q_0 \in [0.991, 1.000]$.
    The values of the binomial test are computed using R \texttt{power.prop.test} function (package \texttt{stats}~\cite{r2024}) with Type~I error $\alphasig = 0.01$ and Type~II error $\betatypetwo = 0.01$. The figure illustrates how shot counts grow rapidly as $q_1$ approaches $q_0$.}
    \label{fig:noise_binomial}
\end{figure}
}

Note that this modelling approach is valid only when the calibrated baseline probability exceeds the target, i.e., $q_{0} > q_{1}$; otherwise, the test cannot reliably distinguish genuine deviations from noise.

In summary, calibrated-noise-baseline methods make inverse and swap tests statistically rigorous under noise, but at the cost of potentially orders of magnitude more shots than QCB suggests. This underscores the need to allocate shot budgets carefully, which we address in \Cref{sec:error_budgeting}.

\section{Budgeting error across program functions}\label{sec:error_budgeting}
Large quantum programs are rarely monolithic: they are composed of many subroutines or functions~\cite{cross2022openqasm,long2023testing,klymenko2025qut}. 
Even if the overall program has a specified fidelity goal, it is not immediately obvious how this tolerance should be distributed across its constituent blocks. 
\Cref{sec:theory} established how the fidelity and the QCB link error probability to the number of required measurements, while \Cref{sec:tests,sec:noise_measurements} demonstrated how these estimates translate into concrete test procedures under both ideal and noisy conditions.

In this section, we extend the per-test analysis to the program level. We begin by introducing the Bures angle~\cite{uhlmann1995geometric}, \cite[Eq. 9.32]{bengtsson2006geometry} as a natural tool for decomposing a global fidelity target into per-function error budgets (\Cref{sec:fidelity_target}). We then show how these per-function targets translate into concrete shot estimates for inverse, swap, and chi-square tests (\Cref{sec:shots_per_block}). Finally, we illustrate the framework with representative examples for programs of varying granularity and hardware-aware weighting schemes (\Cref{sec:shots_per_block_examples}). Together, these results provide practitioners with a methodology for allocating verification resources across complex quantum applications.

\subsection{Derivation of fidelity targets per block/function}\label{sec:fidelity_target}
Suppose we are testing a large quantum program consisting of multiple subroutines.
When decomposing a quantum program into multiple functions (or blocks), it is natural to ask how to distribute the overall error tolerance across the individual components. A convenient way to do so is by using the Bures angle, a metric derived from fidelity~\cite{uhlmann1995geometric}, \cite[Eq. 9.32]{bengtsson2006geometry}. Recall that the Bures angle between two states $\rho$ and $\sigma$ is defined as
$$
A(\rho,\sigma) = \arccos \sqrt{F(\rho,\sigma)}, \quad A(\rho,\sigma) \in [0, \pi/2].
$$
The Bures angle has two key properties that make it attractive for budgeting errors: it is contractive and obeys the triangle inequality~\cite{wu2020quantum}. 

If the program is composed of $k$ functions, and we define hybrid states by replacing the first $j$ functions with their defective and/or noisy versions while keeping the rest ideal, then the triangle inequality yields
$$
A(\rho, \sigma) \leq \sum_{j=1}^k A_j,
$$
where $A_j$ is the Bures angle error contributed by the $j$-th function. 

Suppose that the programmer specifies a program-level fidelity target $F_{\mathrm{prog}}$. This corresponds to an angle budget
\begin{equation}\label{eq:Theta_*}
\Theta_\star = \arccos \sqrt{F_{\mathrm{prog}}}.    
\end{equation}
To guarantee that the program satisfies the fidelity constraint, it suffices to assign per-function angle budgets $\{\theta_j\}$ such that
$$
\sum_{j=1}^k \theta_j \leq \Theta_\star.
$$
Each block then has an individual fidelity target
\begin{equation}\label{eq:f_j_in_terms_of_theta}
F_j^{\text{target}} 
= \cos^2 \theta_j 
\overset{\text{TS at } \theta_j \to 0}{=}
1 - \theta_j^2 + O\left(\theta_j^4\right) \approx 1 - \theta_j^2.    
\end{equation}

In practice, the distribution of the angle budgets can be guided by weights that reflect the relative susceptibility of each block to error. For example, if block $j$ contains $g^{(1)}_j$ one-qubit gates and $g^{(2)}_j$ two-qubit gates with corresponding error rates $r_1$ and $r_2$, one may set
\begin{equation}\label{eq:w_j}
w_j = g^{(1)}_j r_1 + g^{(2)}_j r_2.    
\end{equation}
In general, this formula may have to be altered for the specific architecture and constraints of a particular quantum computer on which the code would be executed, but the general flow of the analysis will hold. For example, if idle errors matter, augment $w_j$ by a depth term $d_j \gamma$, where $d_j$ is the block depth and $\gamma$ the idle error rate per layer. 

The per-block angles are then chosen as
\begin{equation}\label{eq:theta_j}
\theta_j = \frac{w_j}{\sum_{\ell=1}^k w_\ell}  \Theta_\star.    
\end{equation}

This allocation may seem counterintuitive. Blocks with larger weights (e.g., those containing many gates or gates with higher error rates) are assigned a proportionally larger share of the global error budget. This increases their angle budget $\theta_j$, which in turn relaxes their fidelity target and reduces the number of shots required for verification. In contrast, blocks with small weights inherit tight angle budgets, pushing their fidelity targets closer to unity and inflating their shot requirements. While this behaviour may seem paradoxical, it follows directly from the principle of proportional allocation: error-prone blocks are permitted to consume a larger fraction of the global tolerance, whereas simpler blocks must be tested more stringently. 

Proportional allocation is not the only possible policy. In some settings, practitioners may prefer to impose stricter verification on heavier blocks, even at the cost of substantially higher overall shot budgets. Hybrid weighting rules that balance susceptibility to error with the need for tighter guarantees on complex subroutines may also be adopted. 

Up to this point, the derivation is independent of the chosen test. Once $\theta_j$ is known, it can be translated into a fidelity target $F^{\text{target}}_j$ and then substituted into the formulas for the inverse, swap, or chi-square tests from \Cref{sec:tests} as shown below.

\subsection{Computing the number of shots per block/function}\label{sec:shots_per_block}

For the inverse test, once $\theta_j$ is determined, substituting \Cref{eq:f_j_in_terms_of_theta} into \Cref{eq:shot_estimate_real_qc} gives
\begin{equation}\label{eq:N_j}
\begin{split}
N_{j\text{, inverse}}
&\lesssim
\frac{\regimefactor_j \ln \Pe}{ \ln F(\rho, \sigma)} 
= 
\frac{\regimefactor_j \ln \Pe}{ \ln\left( \cos^2 \theta_j \right)} \\
&\overset{\text{TS at } \theta_j \to 0}{=}
-\frac{\regimefactor_j \ln \Pe}{ \theta_j^2} + \frac{\regimefactor_j \ln \Pe}{6} + O\left(\theta_j^2\right)
\approx
-\frac{\regimefactor_j \ln \Pe}{ \theta_j^2},
\end{split}
\end{equation}
where $\regimefactor_j \in [1,2]$ accounts for the pure versus mixed state regimes. This connects program-level fidelity directly to per-block (or per-function) verification costs.

For the swap test, we apply the same principles, substituting \Cref{eq:f_j_in_terms_of_theta} in \Cref{eq:swap_real_bound}
\begin{align*}
N_{j\text{, swap}}
&\lesssim
\frac{\regimefactor_j \ln \Pe}{\ln\left[\frac{1}{2} + \frac{1}{2}F(\rho,\sigma)\right]}
= \frac{\regimefactor_j \ln \Pe}{\ln\left[\frac{1}{2} + \frac{1}{2}\cos^2(\theta_j)\right]} \\
&\overset{\text{TS at } \theta_j \to 0}{=}
-\frac{2\regimefactor_j \ln \Pe}{ \theta_j^2} -\frac{\regimefactor_j \ln \Pe}{6} + O\left(\theta_j^2\right)
\approx
-\frac{2\regimefactor_j \ln \Pe}{ \theta_j^2}.
\end{align*}

For the chi-square test, a closed-form general solution is not available. Instead, we examine two analytically tractable use cases that expose the range of possible shot requirements.
For example, when the discrepancy between the actual and expected distributions is small, the required number of shots satisfies exploring the case of comparing close states; substituting \Cref{eq:f_j_in_terms_of_theta} in \Cref{eq:n_small_discr}
\begin{align*}
N_{j,~\chi^2\text{-small, ideal}} 
& \ge \frac{\lambda(k-1, \alphasig, 1-\betatypetwo)}{8  \left[ 1 - \sqrt{F(\rho, \sigma)} \right]}
= \frac{\lambda(k-1, \alphasig, 1-\betatypetwo)}{8  \left[ 1 - \cos(\theta_j) \right]} \\
&\overset{\text{TS at } \theta_j \to 0}{=}
\frac{\lambda(k-1, \alphasig, 1-\betatypetwo)}{ 4 \theta_j^2} + \frac{\lambda(k-1, \alphasig, 1-\betatypetwo)}{48} + O\left(\theta_j^2\right)
\approx
\frac{\lambda(k-1, \alphasig, 1-\betatypetwo)}{ 4 \theta_j^2}.
\end{align*}
When the measurement basis attains fidelity, using \Cref{eq:N_fidelity_attaining}, the shot requirement is bounded by
\begin{align*}
N_{j,~\chi^2\text{-attaining, ideal}}
& \le \frac{\lambda(k-1, \alphasig, 1-\betatypetwo)}{\frac{1}{4}\left[1-\sqrt{F(\rho, \sigma)}\right]^2}
= \frac{\lambda(k-1, \alphasig, 1-\betatypetwo)}{\frac{1}{4}\left[1-\cos(\theta_j)\right]^2} \\
&\overset{\text{TS at } \theta_j \to 0}{=}
\frac{16 \lambda(k-1, \alphasig, 1-\betatypetwo)}{ \theta_j^4} + \frac{8\lambda(k-1, \alphasig, 1-\betatypetwo)}{3 \theta_j^2} + \frac{11\lambda(k-1, \alphasig, 1-\betatypetwo)}{45} + O\left(\theta_j^2\right) \\
&\approx
\frac{16 \lambda(k-1, \alphasig, 1-\betatypetwo)}{ \theta_j^4}.
\end{align*}

In summary, inverse and swap scale as $N = O(\theta_j^{-2})$, while chi-square (small-discrepancy) also scales as $O(\theta_j^{-2})$. Only the chi-square fidelity-attaining bound grows as $O(\theta_j^{-4})$, highlighting its less practical (as it is an optimistic case) but theoretically important difference.

\subsection{Illustrative examples for inverse test}\label{sec:shots_per_block_examples}

To preserve space, we demonstrate representative examples for the inverse test only; the same process applies to swap and chi-square tests, with swap approximately doubling the cost and chi-square potentially requiring orders of magnitude more shots depending on the case. 

To simplify the notation below, we define $N_j := N_{j\text{, inverse}}$.

\example[Few Functions]{

Suppose the target program fidelity is $F_{\mathrm{prog}}=0.99$, giving as per \Cref{eq:Theta_*}
$$
\Theta_\star = \arccos \sqrt{0.99} \approx 0.100 \ \text{rad}.
$$
Assume the program has three blocks with weights $w=(1,2,3)$. The per-block angle budgets as per \Cref{eq:theta_j} are
$$
\theta_1 \approx 0.017,\qquad \theta_2 \approx 0.033,\qquad \theta_3 \approx 0.050,
$$
summing to $\Theta_\star$. The corresponding fidelity targets as per \Cref{eq:f_j_in_terms_of_theta} are
$$
F_1^{\text{target}} \approx 0.9997,\quad F_2^{\text{target}} \approx 0.9989,\quad F_3^{\text{target}} \approx 0.9975.
$$
With $\regimefactor_j=1$ and acceptance error $\Pe=0.05$, the required shot counts as per \Cref{eq:N_j} are
$$
N_1 \approx 1.1\times 10^4,\quad N_2 \approx 2.7\times 10^3,\quad N_3 \approx 1.2\times 10^3.
$$
Thus, verification is feasible with a few thousand shots per block.
}

\example[Many Functions]{

Keep the same program-level target $F_{\mathrm{prog}}=0.99$ so that $\Theta_\star \approx 0.100$ rad, but now assume the program has $k=10{,}000$ blocks of roughly equal weight. Then each block receives an angle budget
$$
\theta_j = \frac{\Theta_\star}{k} = \frac{\arccos \sqrt{0.99}}{10{,}000} \approx 1.0\times 10^{-5}.
$$
The corresponding fidelity target per block is
$$
F_j^{\text{target}} \approx 0.9999999999.
$$
With $\regimefactor_j=1$ and $\Pe=0.05$, the required number of shots per block is
$$
N_j \approx 3\times 10^{10},
$$
which is impractical. This illustrates how fine-grained decomposition can inflate costs.
}

\example[Gate-Driven Weights]{

To reflect each block’s error exposure, as per \Cref{eq:w_j}, we define weights from one- and two-qubit gate counts and their calibrated error rates:
$$
w_j = g^{(1)}_j r_1 + g^{(2)}_j r_2.
$$

Take $F_{\text{prog}}=0.99$, $\regimefactor_j = 1$, and $\Pe=0.05$. Assume hardware with $r_1=10^{-11}$ (1q) and $r_2=10^{-10}$ (2q). These numbers are based on the Quantinuum estimates~\cite{dasu2025breaking}, where they plan to achieve the logical error rates between $6\times10^{-10}$ and $5\times10^{-14}$ on the actual quantum computers. Consider a program containing 100 functions partitioned into three archetypes\footnote{We can think of an archetype as a representative class of functions (or modules) that share similar structural and behavioural characteristic.}:
\begin{equation*}
\begin{split}
\text{A: }& (g^{(1)},g^{(2)})=(5\times10^4,\ 1\times10^4),\ n_A=10;\\
\text{B: }& (g^{(1)},g^{(2)})=(2\times10^4,\ 4\times10^3),\ n_B=40;\\
\text{C: }& (g^{(1)},g^{(2)})=(5\times10^4,\ 2\times10^4),\ n_C=50;
\end{split}
\end{equation*}
where $n_{(\cdot)}$ is the number of instances of a given function.
The per-instance weights are
$$
w_A=1.5\times10^{-6},\qquad w_B=6.0\times10^{-7},\qquad w_C=2.5\times10^{-6},
$$
and the total weight
$$
W =  n_Aw_A+n_Bw_B+n_Cw_C 
=1.64\times10^{-4}.
$$
Hence the per-instance angle budgets are
$$
\theta_A \approx 9.2\times10^{-4},\quad
\theta_B \approx 3.7\times10^{-4},\quad
\theta_C \approx 1.5\times10^{-3}.
$$
By construction, the per-function budgets satisfy
$\sum_j {n_j \theta_j} = \Theta_\star \approx 0.1$,
so that the aggregate allocation across all function instances recovers the global program budget.

The corresponding fidelity targets per archetype are
$$
F_A^{\text{target}} \approx 0.9999992,\quad
F_B^{\text{target}} \approx 0.9999999,\quad
F_C^{\text{target}} \approx 0.9999977,
$$
and the required shots are
$$
N_A \approx 3.6\times10^{6},\qquad
N_B \approx 2.2\times10^{7}, \qquad
N_C \approx 1.3\times10^{6}.
$$
If one wishes to be maximally conservative for mixed-mixed behaviour, multiply each by at most two ($\regimefactor=2$). 

This example shows how hardware-aware weighting integrates naturally with the Bures angle framework.
}

The examples above illustrate how program-level fidelity goals can be decomposed into concrete shot allocations. We now turn to a broader discussion of the implications, limitations, and future directions of this framework.

\section{Discussion}\label{sec:discussion}

This work developed a principled framework for budgeting measurement shots in quantum program testing. 
Building on theoretical foundations (\Cref{sec:theory}), we analyzed three representative test constructions (inverse, swap, and chi-square) under both idealized and noisy conditions (\Cref{sec:tests,sec:noise_measurements}), and extended the analysis to the program level by introducing Bures-angle-based error partitioning across multiple functions (\Cref{sec:error_budgeting}). 
Here we reflect on the main insights, highlight limitations, and outline avenues for future work.  

\subsection{Summary of results}

\paragraph{Theoretical foundations}  
Using the QCB, fidelity, and trace distance, we established general formulas for relating error probability to shot count. 
In the pure or pure-mixed regimes, closed-form expressions exist, while in the mixed-mixed regime we obtained bounds with the upper bound that differ by at most a factor of two. 
These results provide universal lower and upper limits on the number of shots required, independent of any specific test construction.

\paragraph{Inverse and swap tests}  
Among concrete tests, the inverse test is the most sample-efficient, with swap incurring roughly a factor-of-two overhead because fidelity is encoded indirectly through an ancilla. 
Both tests admit closed-form fidelity-based estimates, are independent of register width, and are susceptible only to Type~II errors in the ideal setting. 

\paragraph{Chi-square test}  
The chi-square test operates directly on measurement distributions, making it easy to implement without modifying circuits. 
However, this convenience comes at the cost of much higher sample requirements~---~often orders of magnitude more than inverse or swap, especially in high-fidelity regimes or when the number of bins grows. 
Moreover, chi-square tests are vulnerable to both Type~I and Type~II errors, and their efficiency depends strongly on how well the readout ``sees'' discrepancies.

\paragraph{Program-level budgeting}  
Using the Bures angle, we showed how a global fidelity goal can be decomposed into per-function fidelity targets and then translated into shot counts. 
This approach highlights a fundamental scaling: per-block verification costs grow\footnote{In the optimistic fidelity attaining scenario for chi-square test, the rate can go up to $N_j = O(\theta_j^{-4})$.} as $N_j = O(\theta_j^{-2})$. 
When the number of functions is small, verification is tractable; when decomposed into thousands of blocks, costs can explode to billions of shots per block, even for modest program-level fidelity goals. 
Weighted distributions (e.g., by gate counts and error rates) provide a more realistic allocation but do not eliminate this scaling challenge.

The scaling we derived for per-function shot counts echoes an important intuition from reliability engineering. In a sequential system, overall reliability is the product of the reliabilities of its components, so the per-component requirement becomes stricter as the system grows~\cite[Sec.~2.2.6]{birolini2010reliability}. Program-level shot budgeting exhibits the same pattern: when the global fidelity goal is partitioned across many functions, each function inherits a smaller angle budget, which inflates the required verification cost. This analogy helps explain why verification becomes impractical when a program is decomposed too finely.

\paragraph{Noise}  
Our analysis revealed that handling noise requires going beyond the simple QCB picture. 
One pragmatic strategy is to treat noise phenomenologically via the parameter $\regimefactor \in [1,2]$, which interpolates between pure and mixed regimes and provides conservative bounds on shot counts. 
A more rigorous strategy calibrates a noise-only baseline and applies a binomial hypothesis test to separate genuine state deviations from random fluctuations. 
This baseline approach enforces explicit control over both Type~I and Type~II errors, but can inflate shot requirements by orders of magnitude compared to $\regimefactor$-based estimates. 
The contrast highlights a key trade-off: $\regimefactor$ offers a convenient rule-of-thumb, while binomial calibration yields stronger guarantees at substantially higher cost. 
We return to this trade-off in Section~\ref{sec:limitations}, where we discuss implications for noise-aware test design.

\subsection{Limitations and future directions}\label{sec:limitations}
Several limitations remain, they also serve as a starting point for several avenues for further research.

\paragraph{Noise modelling} Our treatment of inverse and swap tests incorporated baseline calibration, yet the optimal and systematic method for constructing such baselines remains underexplored. For chi-square and other statistical tests, noise-aware formulations are also open challenges. In particular, extending hypothesis testing frameworks to account for drift, correlated errors, and device-dependent baselines represents an important direction for future research. 

\paragraph{Property-based testing.} All derivations assumed knowledge of the expected distribution or state. In practice, developers may wish to verify structural properties (e.g., symmetry~\cite{laborde2023testing} or conservation laws~\cite{zhan2024learning}) rather than exact outcomes. Adapting shot-budget analysis to such property testing remains an open problem. 

\paragraph{Static vs. dynamic analysis vs. state vector} Our results assume repeated execution on hardware. For many functions, dynamic testing may be prohibitive, motivating hybrid approaches that leverage static analysis (see~\cite{murillo2024challenges,ramalho2024testing} for a review) or state vector simulation for smaller subcircuits~\cite{miranskyy2025feasibility,ye2025measurement}.

\paragraph{Multiple dimensions of cost} Although this paper has focused primarily on shot counts, they represent only one axis of verification cost. Other dimensions include the complexity of constructing the required circuits~\cite[Tbl.~2]{miranskyy2025feasibility}, the overhead of additional ancilla qubits, and the practical effort of transpilation and compilation. These dimensions can dominate in real-world settings, meaning that test selection should balance both sampling efficiency and implementation effort. Thus, integrating circuit complexity (gate counts, ancilla overhead, transpilation cost) with shot-budget analysis to guide practitioners in choosing appropriate tests is a good avenue of future research.  

\paragraph{Tool support} We provide sample code for computing the budget-related formulas. However, automating budget allocations and per-block shot planning within quantum software engineering toolchains, enabling developers to estimate verification costs before running large-scale experiments is a good future task. Moreover, in future toolchains, compilers could generate inverse or swap circuits on the fly, enabling fidelity-based comparison as a built-in feature. 

\section{Conclusions}\label{sec:conclusion}
This work established a unified framework for estimating the number of measurements required to verify quantum programs. We began with theoretical bounds based on the quantum Chernoff bound, fidelity, and trace distance, and then translated these into concrete shot estimates for inverse, swap, and chi-square tests under both idealized and noisy conditions. Our analysis confirmed that the inverse test is most sample-efficient, the swap test incurs roughly a factor-of-two overhead, and chi-square tests, while circuit-light, are typically orders of magnitude more demanding.

Extending beyond individual tests, we introduced a Bures-angle-based method for distributing program-level fidelity budgets across subroutines, showing how fine-grained decomposition can render verification intractable. Together, these results provide actionable guidance for practitioners in planning verification strategies and allocating shot budgets.

\section*{Acknowledgements}

All data manipulations and figures are generated using R packages \texttt{tidyverse}~v.2.0.0~\cite{wickham2019tidyverse} and \texttt{ggplot2}~v.3.5.2~\cite{wickham2016ggplot2}.
Some symbolic calculations were verified using Mathematica~v.14.3~\cite{Mathematica}.
An initial version of the demo code was generated using ChatGPT~5~\cite{chatgpt5} and subsequently refined by the author.

\printbibliography

\appendix
\section{Upper and lower bounds of $Q$ for mixed states}\label{sec:q_mixed_boundaries}
\subsection{Lower bound of $Q$}
From \cite[Eq.~12]{audenaert2007preprint}, the following relation holds:
\begin{equation}\label{eq:q_and_t}
Q(\rho,\sigma) + T(\rho,\sigma) \ge 1.
\end{equation}
Moreover, \cite[Eq.~41]{fuchs1999cryptographic} establishes the connection between trace distance and fidelity:
\begin{equation}\label{eq:t_and_f}
1 - \sqrt{F(\rho,\sigma)} \le T(\rho,\sigma) \le \sqrt{1-F(\rho,\sigma)}.
\end{equation}

To obtain a lower bound for $Q(\rho,\sigma)$, we minimize it using \Cref{eq:q_and_t}. Since this requires maximizing $T(\rho,\sigma)$, we take the upper bound from \Cref{eq:t_and_f}. Substituting gives
\begin{equation*}
\tcboxmath[colback=white,colframe=gray]{
Q(\rho,\sigma) \ge 1 - T(\rho,\sigma) \ge 1 - \sqrt{1-F(\rho,\sigma)}.
}
\end{equation*}
 
\subsection{Upper bound of $Q$}
As shown in \cite[Eq.~13]{audenaert2007preprint}, the $Q(\rho,\sigma)$ admits the following upper bound:
$$
Q(\rho,\sigma) \le \Tr\left[\rho^{1/2}\sigma^{1/2}\right]
= \left\| \rho^{1/4}\sigma^{1/2}\rho^{1/4} \right\|_{1}
\le \left\| \rho^{1/2}\sigma^{1/2} \right\|_{1}.
$$
Thus, based on the definition of fidelity in \Cref{eq:fidelity},
\begin{equation*}
\tcboxmath[colback=white,colframe=gray]{
Q(\rho,\sigma) \le \sqrt{F(\rho,\sigma)}.
}
\end{equation*}

\section{Compute the number of shots in terms of the trace distance}\label{sec:trace_distance}
\subsection{Trace distance}
While fidelity provides one way to quantify the similarity of quantum states, another widely used measure is the trace distance. For two density matrices (quantum states) $\rho$ and $\sigma$, the trace distance is defined as
$$
T(\rho, \sigma) = \frac{1}{2}  \|\rho - \sigma\|_1, \quad T(\rho, \sigma) \in [0, 1].
$$
The trace distance satisfies $T(\rho,\sigma) = 0$ if and only if $\rho = \sigma$ (the states are identical), and $T(\rho,\sigma) = 1$ if and only if $\rho$ and $\sigma$ have orthogonal supports (perfectly distinguishable).  Thus, $T(\rho, \sigma)$ ranges between $0$ and $1$, with smaller values indicating greater similarity.

Using known inequalities from~\cite[Sec. 9.2.3]{nielsen_chuang_2010}, we can reformulate the shot count estimates from \Cref{sec:fidelity} in terms of $T$ rather than $F$.

\paragraph{Pure-pure case}
From~\cite[Eq. 9.99]{nielsen_chuang_2010}, the relationship between trace distance and fidelity for two pure states is:
$$
    T(\rho, \sigma) = \sqrt{1 - F(\rho, \sigma)} \quad \Rightarrow \quad F(\rho, \sigma) = 1 - T(\rho, \sigma)^2.
$$
Substituting it into \Cref{eq:shot_estimate_pure} gives
$$
\tcboxmath[colback=white,colframe=gray]{
N_\text{pure} \sim \frac{\ln \Pe}{\ln \left[1-T(\rho, \sigma)^2\right]}.
}
$$
Thus, $N_\text{pure}$ can be written purely in terms of trace distance.

\paragraph{Pure-mixed case}
When one state is pure and the other is mixed, \cite[Eqs. 9.110 and 9.111]{nielsen_chuang_2010} give:
$$
    1 - F(\rho, \sigma) \le T(\rho, \sigma) \le \sqrt{1-F(\rho, \sigma)}.
$$
Inverting these inequalities provides bounds on fidelity in terms of trace distance:
$$
    1 - T(\rho, \sigma) \le F(\rho, \sigma) \le 1-T(\rho, \sigma)^2.
$$
Substituting into \Cref{eq:shot_estimate_pure} gives corresponding bounds for the shot count:
\begin{equation*}
\tcboxmath[colback=white,colframe=gray]{
\frac{\ln \Pe}{\ln \left[1-T(\rho, \sigma)\right]}
\lesssim
N_\text{pure-mixed}
\lesssim
\frac{\ln \Pe}{\ln  \left[1-T(\rho, \sigma)^2\right]}.
}
\end{equation*}
Thus, in the pure-mixed scenario, the required number of shots falls between these two limits.

\paragraph{Mixed-mixed case}
For two mixed states, the relationship between trace distance and fidelity is bounded \cite[Eq. 9.110]{nielsen_chuang_2010}:
$$
    1 - \sqrt{F(\rho, \sigma)} \le T(\rho, \sigma) \le \sqrt{1-F(\rho, \sigma)}.
$$
Solving for fidelity yields:
$$
    \left[1 - T(\rho, \sigma)\right]^2 \le F(\rho, \sigma) \le 1-T(\rho, \sigma)^2
$$
Substituting these bounds into \Cref{eq:shot_estimate_mixed} gives corresponding estimates for the number of shots in the mixed-mixed case:
\begin{equation*}
\tcboxmath[colback=white,colframe=gray]{
\frac{\ln \Pe}{\ln \left[1 - \sqrt{1-\left[1 - T(\rho, \sigma)\right]^2}\right]}
= 
\frac{\ln \Pe}{\ln \left[1 - \sqrt{2 T(\rho, \sigma) - T(\rho, \sigma)^2}\right]}
\lesssim
N_\text{mixed}
\lesssim
\frac{2\ln \Pe}{\ln \left[ 1-T(\rho, \sigma)^2\right]}.
}
\end{equation*}
Here again, the lower bound may be considerably smaller than in the $N_\text{pure}$ and $N_\text{pure-mixed}$ cases, while the upper bound can be up to twice as large.

\subsubsection{Comparison of $N_\text{pure}$, $N_\text{pure-mixed}$,  and $N_\text{mixed}$}
\begin{figure}[tbh]
    \centering
    \includegraphics[width=1.0\linewidth]{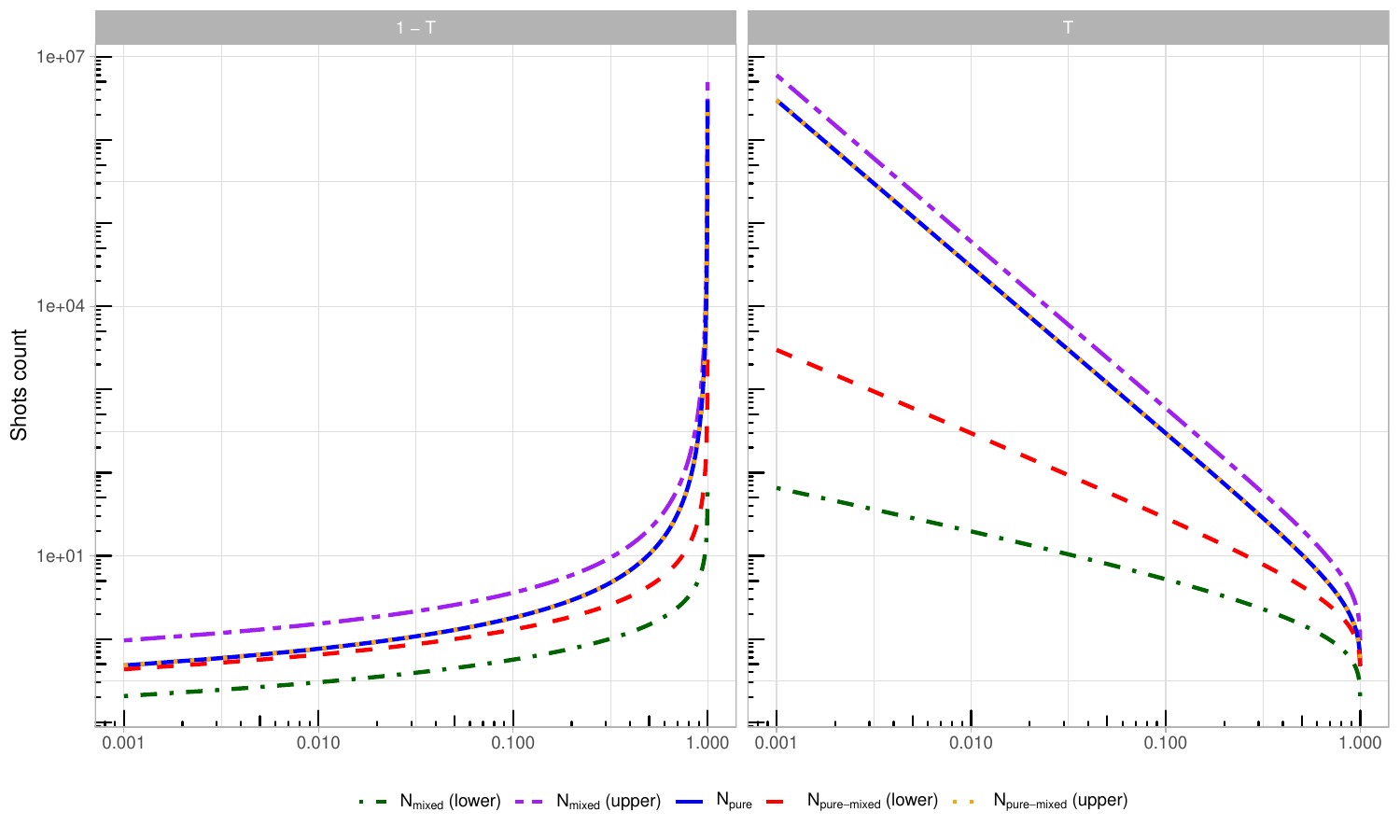}
    \caption{
    The number of measurement shots $N$ required to achieve error probability $\Pe = 0.05$ is shown as a function of the trace distance $T \in [0.001, 0.999]$ (right pane). The curves depict the pure-state case $N_{\text{pure}}$, as well as the lower and upper bounds for the pure-mixed case $N_{\text{pure-mixed}}$ and the mixed-state case $N_{\text{mixed}}$. As $T \to 0$, the required number of shots diverges exponentially; therefore, the left pane shows the same data plotted against $1-T$ for improved readability. Note that the upper bound of $N_{\text{pure-mixed}}$ coincides with $N_{\text{pure}}$.}
    \label{fig:tr_f_vs_n}
\end{figure}

\Cref{fig:tr_f_vs_n}  illustrates how the required number of measurement shots depends on trace distance. Conceptually, the dynamics are similar to those observed for fidelity (\Cref{fig:f_vs_n}), but with inverted behaviour: while fidelity diverges as it approaches 1, the trace distance diverges as it approaches 0. The difference is that, for fidelity, the pure and pure-mixed cases coincide, whereas for trace distance the upper boundary of the pure-mixed case coincides with the pure case, but unlike fidelity, the pure-mixed case also has a distinct lower boundary.

\section{Probability of observing zero-string in the inverse test}\label{sec:inverse_test_zero_string_probability}
Recall the construction of the inverse test as in~\cite[Sec. 3-B4]{miranskyy2025feasibility}. The circuit under test $U$ is applied to the input state $\ket{\psi_I}$, producing the actual state
$$
\ket{\psi_A} = U \ket{\psi_I}.
$$

Let $\ket{\psi_E}$ be the expected state of the circuit. Choose a unitary $Z$ such that
\begin{equation}\label{eq:z_and_E}
Z \ket{\psi_E} = \ket{0^n}.
\end{equation}
Applying $Z$ to the actual state yields
\begin{equation}\label{eq:psi_R}
    \ket{\psi_R} = Z \ket{\psi_A}.
\end{equation}

Measuring all $n$ qubits of $\ket{\psi_R}$ in the computational basis, the probability of obtaining $0^n$ is
$$
P(M_{\ket{\psi_R}} = 0^n) = \left | \braket{0^n | \psi_R} \right |^2.
$$
Substituting \Cref{eq:psi_R} gives
\begin{equation}\label{eq:p_m_intermediate}
P(M_{\ket{\psi_R}} = 0^n) = \left | \braket{0^n | Z | \psi_A} \right |^2.
\end{equation}

Left-multiplying \Cref{eq:z_and_E} by $Z^\dagger$ and using the rule $Z^\dagger Z = ZZ^\dagger = I$ gives
$$
Z^\dagger Z  \ket{\psi_E} = Z^\dagger  \ket{0^n}
\quad \Rightarrow \quad
\ket{\psi_E} = Z^\dagger  \ket{0^n}.
$$
Taking the adjoint of the whole equation yields\footnote{By using the rules  $(AB)^\dagger=B^\dagger A^\dagger$, $(\bra{v})^\dagger=\ket{v}$, $(\ket{v})^\dagger=\bra{v}$,  and $(Z^\dagger)^\dagger=Z$.}
$$
\left(\ket{\psi_E} \right)^\dagger = \left(Z^\dagger  \ket{0^n}\right)^\dagger
\quad \Rightarrow \quad
\bra{\psi_E} = \bra{0^n} Z.
$$
Substituting this into \Cref{eq:p_m_intermediate} results in
\begin{equation*}
\tcboxmath[colback=white,colframe=gray]{
P(M_{\ket{\psi_R}} = 0^n) = \left | \braket{\psi_E | \psi_A} \right |^2,
}
\end{equation*}
which equals the fidelity between the two pure states $\ket{\psi_E}$ and $\ket{\psi_A}$.

\section{Quantum Chernoff bound for the swap test}\label{sec:qcb_for_swap}
The expected state of the auxiliary qubit is the pure state
$$
\sigma_a
= 
\ket{0}\bra{0}
= 
\begin{bmatrix}
1 & 0 \\
0 & 0
\end{bmatrix}.
$$
As shown in~\cite[p.~167902-2]{buhrman2001}, the measurement probabilities for the swap test are
$$
P(M_{q_a} = 0) =  \frac{1}{2}+\frac{1}{2} F(\rho, \sigma),
$$
$$
P(M_{q_a} = 1) =  \frac{1}{2}-\frac{1}{2} F(\rho, \sigma).
$$

Thus, the reduced density matrix of the auxiliary qubit is
$$
\rho_a
= 
\begin{bmatrix}
\frac{1+F(\rho, \sigma)}{2} & \cdot \\
\cdot & \frac{1-F(\rho, \sigma)}{2}
\end{bmatrix}.
$$

Since one of the two states is pure, by \Cref{eq:q_pure} the QCB simplifies to
\begin{equation*}
\tcboxmath[colback=white,colframe=gray]{
Q(\rho, \sigma) = \Tr(\rho_a \sigma_a) = \frac{1+F(\rho, \sigma)}{2}.
}
\end{equation*}

\end{document}